\documentclass[9pt,twocolumn,twoside]{pnas-new}
\templatetype{pnasresearcharticle} 
\title{Rototaxis: localization of active motion under rotation}
\author[a,b]{Yuanjian Zheng}
\author[b]{Hartmut L\"{o}wen} 
\affil[a]{Division of Physics and Applied Physics, School of Physical and Mathematical Sciences, Nanyang Technological University, Singapore 637371, Singapore}
\affil[b]{Institut f\"{u}r Theoretische Physik II: Weiche Materie, Heinrich-Heine-Universit\"{a}t D\"{u}sseldorf, D-40225 D\"{u}sseldorf, Germany}
\leadauthor{Zheng} 
\significancestatement{The ability of living organisms to move in a favorable direction in response to the presence of food or danger, is critical to survival and the evolutionary success of its species. We introduce a new mechanism of locomotion in a rotating environment –  rototaxis, where an object reacts to a change in the speed at which the environment around it is moving. This mechanism explains how living organisms or artificial particles can survive in rotating environments such as swirls by presenting two complementary modes of motion with contrasting survival or evolutionary goals. A living organism that positions itself to facilitate the ease of switching between these two modes of movement may be the most optimally primed for survival.}
\authorcontributions{Author contributions: Y.Z. and H.L. designed, performed research and wrote the paper.}
\authordeclaration{The authors declare no competing interest.}
\correspondingauthor{\textsuperscript{1}To whom correspondence should be addressed. E-mail: zhen0117@e.ntu.edu.sg}

\begin{abstract}
	The ability to navigate in complex, inhomogeneous environments is fundamental to survival at all length scales, giving rise to the rapid development of various subfields in bio-locomotion such as the well established concept of chemotaxis. In this work, we extend this existing notion of taxis to rotating environments and introduce the idea of  ``roto-taxis" to bio-locomotion. In particular, we explore both overdamped and inertial dynamics of a model synthetic self-propelled particle in the presence of constant global rotation, focusing on the particle's ability to localize near a rotation center as a survival strategy. We find that in the overdamped regime, the swimmer is in general able to generate a self restoring active torque that enables it to remain on stable epicyclical-like trajectories. On the other hand, for underdamped motion with inertial effects, the intricate competition between self-propulsion and inertial forces, in conjunction with the rototactic torque leads to complex dynamical behavior with non-trivial phase space of initial conditions which we reveal by numerical simulations. Our results are relevant for a wide range of setups, from vibrated granular matter on turntables to microorganisms or animals swimming near swirls or vortices.	
\end{abstract}

\dates{This manuscript was compiled on \today}
\doi{\url{www.pnas.org/cgi/doi/10.1073/pnas.XXXXXXXXXX}}

\begin{document}

\maketitle
\thispagestyle{firststyle}
\ifthenelse{\boolean{shortarticle}}{\ifthenelse{\boolean{singlecolumn}}{\abscontentformatted}{\abscontent}}{}
	
Microorganisms and artificial microswimmers alike  \cite{ElgetiGompper2015,BechingerVolpe 2016,MarchettiSimha2013}, are highly capable of adapting their motion in response to gradients present in external stimuli. This inherent property of life, often collectively referred to as {\it taxis} \cite{Dusenbery2009}, is not only extremely crucial to survival and the emergent evolutionary fitness of living species, but also underpins the complex dynamics of artificial self-propelled colloidal particles, that are showing increasing promise for a wide range of industrial and technological applications \cite{Ebbens2016}.

Chemotaxis \cite{Eisenbach2004}, as one of the most prominent and established forms of taxis, is the ability of an organism (or a particle) to steer its motion internally towards spatial regions of higher or lower concentration of one or more chemical species. More specifically, chemo-attraction signaling or positive chemotaxis, is the underlying drive of an organism towards higher chemical concentration and is thus often a favorable strategy to adopt in the presence of nutrients or fuel sources. On the other hand, chemo-repulsion (negative chemotaxis) is desired if the chemical species present are instead for instance, toxins. In addition to bio-locomotion, chemotaxis also plays an important role in the behavior of a large class of synthetic colloidal microswimmers \cite{HongVelegol2007, PohlStark2014, SahaRamaswamy2014, LiebchenCates2015}, governing dynamical pattern formation in active many-body systems \cite{PohlStark2014, SahaRamaswamy2014, LiebchenCates2015}. 

Tactic responses are also known for many other physical stimuli such as variation in the intensity of light (phototaxis) \cite{JekleyArendt2008, BennettGolestanian2015, GiomettoStocker2015, LozanoBechinger2016, DaiTang2016, GarciaPeyla2013, PalagiFischer2019,MoysesGrier2016}, external magnetic (magnetotaxis) \cite{KlumppFaivre2016, RupprechtBocquet2016, WaisbordCottinBitzonne2015,VincentiClement2019} or gravitational field strengths (gravitaxis) \cite{RichterHaeder2007, tenHagenbechinger2014, CampbellEbbens2013, WoffStark2013}, temperature (thermotaxis) \cite{BahatEisenbach, BregullaCichos2016}, viscosity (viscotaxis) \cite{LiebchenLoewen2018}, orientation of the solvent flow field (rheotaxis) \cite{FuStocker2012, MathijssenZoettl2019, BrosseauShelley2019} and even topographical gradients that may be present in distorted solids or environments (topotaxis) \cite{ParkLevchenko2016,ParkLevchenko2018,SchakenraadGiomi2019}.

In this spirit, we consider a form of taxis where the external stimulus takes the form of a {\it gradient in the local kinetic energy}. This form of taxis has several advantages as a survival strategy in a rotating environment \cite{TaramaLoewen2014,KuechlerLoewen2016,SupekarDunkel2019}. Firstly, a propensity for the positive gradient which we term positive {\it rototaxis}, corresponding to the drive towards a region where solvent flow is large, would be an ideal strategy for enhanced transport in situations where an organism seeks to escape from death or destruction in vortex environments such as whirlpools, maelstroms, and industrial or laboratory centrifuges \cite{PetersonBusscher2012,LiuRunge2017}, or where it is attempting to chase down a prey \cite{AkanyetiLiao2017}. Conversely, negative rototaxis results in the localization of dynamical trajectories in the vicinity of rotation centers and would thus be largely favorable for remaining near static food sources, maintaining proximity to swarm centers \cite{ViscekZafeiris2012}, or simply energy saving by not swimming against the flow in low vorticity scenarios. In this context, we note that ideal rototaxis occurs when the swim orientation perfectly aligns along the gradient of kinetic energy $\nabla \vec{u}^2(\vec {r})$ where $\vec{u}(\vec{r})$ is the flow velocity. This should be contrasted to ordinary rheotaxis \cite{FuStocker2012,MathijssenZoettl2019,BrosseauShelley2019} where the alignment of the swim axis is along the flow direction ${\vec u}(\vec {r})$ itself.  

However, despite its clear survival advantages as a taxis strategy, generating negative rototaxis is a considerably difficult task for microswimmers. This is due to the centrifugal force in the rotating frame of the system that tends to expel the particle or organism away from its rotation center. In this work, we introduce a model 2D microswimmer capable of generating a tunable active torque that sustains both negative and positive rototaxis by simply altering its initial swim orientation with respect to its position in the flow field. This is made possible in the presence of {\it inertia} that is increasingly being recognized to play a pivotal role in the dynamics of microswimmers \cite{ScholzLoewen2018,FouxonOr2019, GiomiMahadevan2013, AttansiViale2014}. 

The paper is organized as follows. In sec. \ref{sec:model}, we model the basic physics and equations of motion for an active microswimmer in a rotating, noise-free environment. We then solve these equations in the overdamped regime in sec. \ref{sec:overdamped} and argue that negative rototaxis is always recovered in the overdamped limit in the presence of any arbitrarily small but finite noise. Within a linear stability analysis, we show that this results in stable epicyclic-like trajectories and rosettes in both the rest and rotating frames of real space. We then numerically solve the non-linear inertial problem in sec. \ref{sec:inertia} to reveal the existence of both positive and negative rototactic phases that is intricately dependent on initial conditions. In addition, we also find a further differentiation in the negative rototactic phase; the emergence of circular periodic limit cycles in addition to the existing epicyclic and rosette solutions found in the overdamped limit. Lastly in sections \ref{sec:discussion} and \ref{sec:conclusion} , we discuss the implications of our work especially in context of survival strategies and recent experimental work in bacterial dynamics under shear flow and draw our conclusions of rototaxis as a general guiding principle of survival for biological microswimmers in rotating environments.

\section{Model}
\label{sec:model}

We begin by considering a single particle consisting of two beads that are point masses, $m_1$ and $m_2$ (i.e. the total mass of the particle is $m=m_1+m_2$) at positions $\vec{r}_1$ and $\vec{r}_2$ connected by a light inextensible rod of length $ l \equiv \vert \vec{r}_1 -\vec{r}_2 \vert$. The particle is confined to a 2D plane and is acting under a global rotation of angular frequency $\vec{\omega}$ about some fixed origin in the plane, corresponding to an effective flow field ${\vec u}({\vec r}) = {\vec \omega}  \times {\vec R}$.

This minimal bead model is designed to capture the essential features of a large class of rod-shaped axial propelled bacteria such as well-studied model organisms {\it Escherichia coli} and {\it Bacillus subtilis} \cite{TaylorSalama2019}, while still allowing for modular generalizations by attachment or removal of additional beads or structures for more complex geometries and swim orientation. Moreover, we note that synthetic experimental realizations of such bead models \cite{NajafiGolestanian2004,KuechlerLoewen2016,RizviMisbah2018a,SukhovHarting2019,GrosjeanVandewalle2019} in the overdamped limit has also recently been achieved \cite{GrosjeanVandewalle2016}.

Activity is introduced to the model microswimmer by a persistent self-propulsion velocity $v_0 \hat{n} $ projecting from the geometrical center of the two beads $\vec{R}$ in a direction along the long axis of the particle towards $\vec{r}_2$ such that each bead is always $l/2$ away from $\vec{R}$, where $\hat{n} \equiv (\vec{r}_2-\vec{r}_1)/l= \left[ \cos \phi, \sin \phi \right]^{T}$ and $\phi(t)$ is the swim orientation phase angle, defined with respect to the rest frame of the system from some positive $x$-axis. 

Frictional effects arising from translational ($\gamma$) and rotational ($\gamma_R$) viscosities, are assumed to be acting on the geometrical center $\vec{R}$ and orientational $\phi$ degrees of freedom, and is also assumed to be linearly proportional to the translational and angular velocities in the rotating frame respectively. The translational and swim orientational equations of motion (in the rest frame) are thus written as
\begin{align}
m \ddot{\vec{R}}+\gamma(\dot{\vec{R}}-\vec{\omega} \times \vec{R}) =\gamma v_0 \hat{n} + \vec{f}(t) \label{eq:eom}
\\ 
J \ddot{\phi}+\gamma_{R}(\dot{\phi}- \omega) = M + \tau(t) + g(t) \label{eq:eom2}
\end{align}
where the translational degree of freedom is the geometrical center $ \vec{R}=\vec{r}_i + \frac{(3-2i)l}{2} \hat{n}$ for $(i=1,2)$ and where $J= \sum_i m_i (\vec{R}-\vec{r}_i)^2= \frac{m}{4} l^2$ is the moment of inertia of the dumbbell-like particle and $\tau (t)$ which controls rototaxis, and is therefore referred to as the {\it rototactic torque}. 

Here this rototactic torque is generated by inertia. To calculate it, the inertial torque $\tau'(t)$ experienced internally by the particle in the rotating frame is identical to $\tau(t)$ in the rest frame (i.e $\tau'=\tau$), and originates from fictitious centrifugal $ \vec{F}^c_i \equiv - m_i \vec{\omega} \times (\vec{\omega} \times \vec{r_i}')$ and Coriolis $-2m \vec{\omega} \times \vec{v}'_i$ components of inertial forces acting individually on each bead with $\vec{v}'_i= \dot{\vec{r}}_i-\vec{\omega} \times \vec{r}_i$ and $X'$ denoting a quantity in the rotating frame. For large damping $\gamma$, eq. [\ref{eq:eom}] leads to small velocities $\vec{v}'_i$ such that the centrifugal component typically dominates over the Coriolis term. Hence the torque $\tau(t)$ in the rest frame can be written as
\begin{align}
\tau(t) &= \sum^2_i (\vec{R} -\vec{r}_i) \times \left( \vec{r}_i m_i \omega^2 \right)  &= -  \frac{m}{2} \tilde{\mu} l \omega^2 (\hat{n} \times \vec{R}) 
\\ &= - \frac{m}{2} \tilde{\mu} l \omega^2  R \sin (\theta - \phi)
\label{eq:torque}
\end{align}
\begin{figure}[tp!]
	\includegraphics[width=\columnwidth]{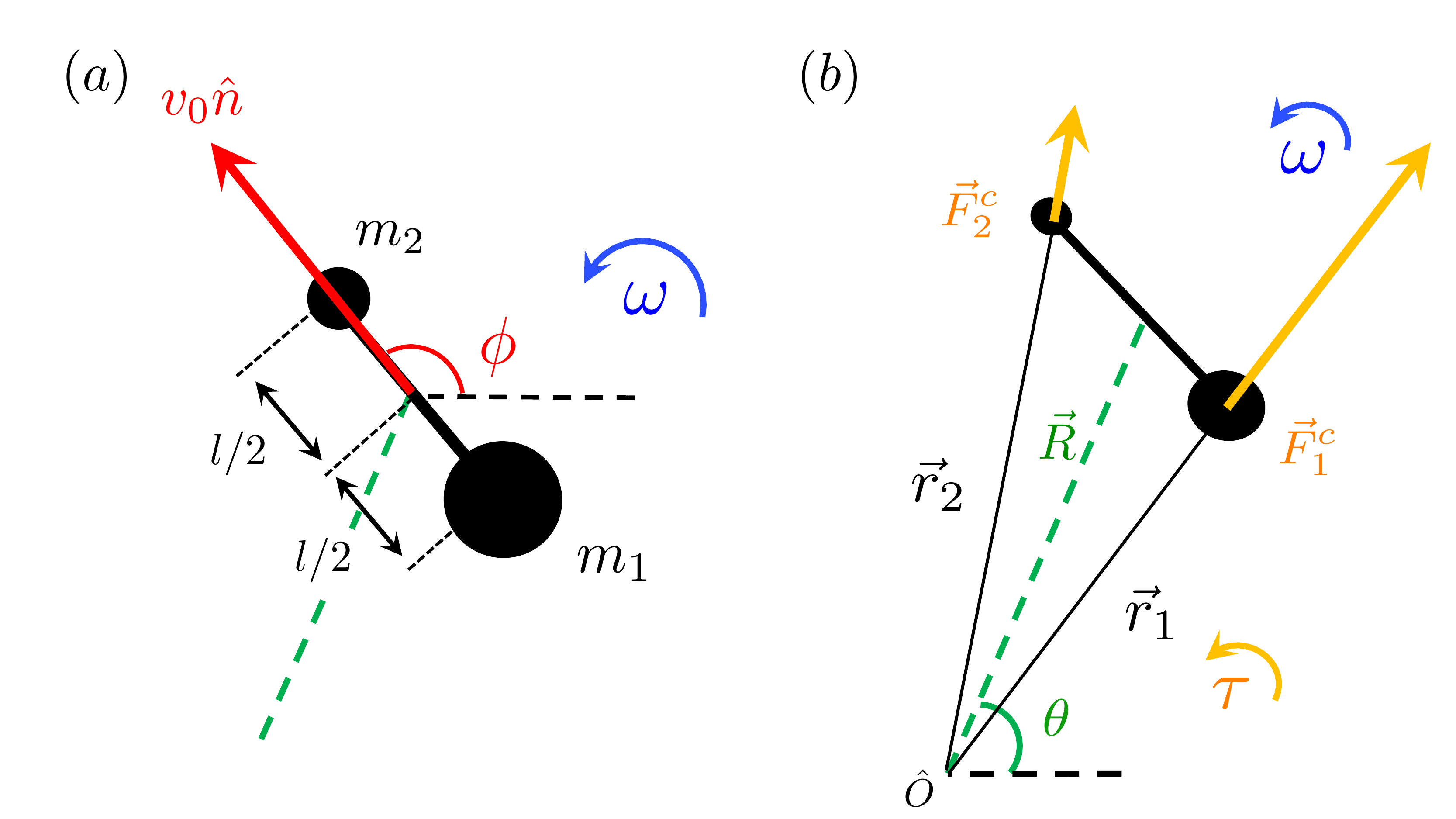}
	\caption{Schematics of the rototactic microswimmer. (a) self-propulsion (red) with swim orientation $\phi$ of an axial swimmer under a constant global rotation $\omega$ (blue) (b) Positional degrees of freedom $R$ and $\theta$ defined with respect to the same axis as $\phi$ from the center of rotation $\hat{O}$.  Torque (orange) generated from centrifugal forces $\vec{F}^c_i$ that may in general be opposing the global rotation $\omega$.}
	\label{fig:schematic}
\end{figure}
where $\tilde{\mu} \equiv \frac{m_1-m_2}{m}$ is the scaled mass difference between the two beads, such that maximal rototactic torque is recovered in the limit of $ \vert \tilde{\mu} \vert \to 1$. Note that here we have parameterized the geometrical center $\vec{R}$ in the polar coordinates where $\theta$ is the positional phase angle of $\vec{R}$, (i.e $\vec{R}= R [\cos \theta, \sin \theta ]^T$) in the rest frame and is defined from the same axis that defines $\phi$. See Fig. \ref{fig:schematic}(a,b) for schematics of the model microswimmer in the rotating environment. 

Here we remark that a torque $\mathcal{T} \sim R \sin \left( \theta - \phi \right)$ that plays the same essential role as $\tau$ in orientating the swimmer towards the rotation center $\hat{O}$, can also be realized by placing a light source at the rotation center for phototactic particles
\cite{JekleyArendt2008, BennettGolestanian2015, GiomettoStocker2015, LozanoBechinger2016, DaiTang2016, GarciaPeyla2013, PalagiFischer2019,MoysesGrier2016} or by placing a pole of a magnetic field at $\hat{O}$ for magnetotactic swimmers \cite{KlumppFaivre2016, RupprechtBocquet2016, WaisbordCottinBitzonne2015, VincentiClement2019}.

For full generality and completeness, we also include contributions from fluctuating forces $\vec{f}(t)$ and torques $\vec{g}(t)$ from the solvent that are typically modelled as Gaussian white noise acting on $\vec{R}$ and $\phi$ respectively \cite{BechingerVolpe 2016}, as well as provision for an internal torque $M$ that allows for the modelling of circle swimmers \cite{KummelBechinger2013}. However for simplicity, $f(t)$, $g(t)$ and $M$ are all assumed to be zero and we consider the scenario of a bottom heavy particle ($m_1 > m_2$) where the rototactic effect is maximal, such that $\tilde{\mu} \to 1$ in our model of rototaxis hereinafter.

\section{Overdamped Dynamics}
\label{sec:overdamped}

In this section, we consider overdamped dynamics of the rototactic particle by neglecting inertial terms ($J \ddot{\phi}=0$ and $m\ddot{\vec{R}}=0$) and any internal torque ($M=0$) in the noise-free limit $(\vec{f}(t)=\vec{g}(t)=0)$ of the rest frame eqs. [\ref{eq:eom}] and [\ref{eq:eom2}]:
\begin{align}
\dot{\vec{R}}&= v_0 \hat{n}  + \vec{\omega} \times \vec{R} \label{eq:eom_o}
\\ 
\dot{\phi} &=- \frac{m}{2} \tilde{\mu} l \omega^2 R \sin (\theta - \phi) / \gamma_{R} + \omega  \label{eq:eom_o2}
\end{align}
These equations include the crucial presence of the rototactic torque $\tau (t)$ that consistently vanishes in the limit of $l \to 0$. and directly corresponds to the limit of strong activity where $v_0 \gg l \omega$ (see Appendix \ref{sec:app_2}).

We note that in the absence of rotation $\vec{\omega}$, the overdamped limit of this model has recently been investigated for a mathematically equivalent phototactic torque $\mathcal{T}$ in a system of phototactically driven Janus particles and found experimentally to result in non-trivial trochoidal trajectories \cite{MoysesGrier2016}, which we subsequently also recover in this section. On the other hand, the zero-torque limit of this model has been studied in \cite{Loewen2019a}.

Now, the radial and angular components of [\ref{eq:eom_o}] explicitly reads 
\begin{align}
\dot{R}&=v_0 \cos (\theta-\phi)\label{eq:eom_oo}
\\
\dot{\theta}&=\omega-v_0 \sin(\theta - \phi)/R\label{eq:eom_oo2}
\end{align}
By introducing the parameter $\kappa_0\equiv \frac{m}{2} \tilde{\mu} l \omega^2/\gamma_R$ and a variable corresponding to the angular difference between positional and orientational phases 
\begin{equation}
\alpha  \equiv \theta - \phi 
\end{equation}
[\ref{eq:eom_o}] and [\ref{eq:eom_o2}] reduces to a system consisting of only two degrees of freedom in $\alpha$ and $R$.
\begin{align}
\dot{R}&=v_0 \cos \alpha \label{eq:eom_alpha}
\\
\dot{\alpha}&= \left( \kappa_0 R-\frac{v_0}{R}\right) \sin \alpha \label{eq:eom_alpha2}
\end{align}
which in turn, is equivalent to a single separable first order differential equation that upon solving yields the following transcendental equation.
\begin{equation}
\frac{R \sin \alpha}{R_{int} \sin \alpha_{int}}=\exp  \left({\frac{\kappa_0(R^2-R^2_{int})}{2v_0}}\right)
\label{eq:transcedental}
\end{equation}
where $R_{int}$ and $\alpha_{int}$ denote initial conditions of $R$ and $\alpha$ at time $t=0$ respectively. 

This system of equations involving $\alpha$ has two fixed points $ \left( R_o^{\star}, \alpha_{o,\pm}^{\star} \right) =  \left( \sqrt{v_0/\kappa_0} , \pm \pi/2 \right)$ corresponding to closed circular trajectories in the $ R$-$\theta$ plane, and can also be seen as the set of solutions where $\dot{\theta}=\dot{\phi}$ in eqs. [\ref{eq:eom_oo}] and [\ref{eq:eom_oo2}], for which $\dot{R}=0$ can only be satisfied at a critical radial distance $R_{o}^{\star}$.

A linear perturbative expansion of [\ref{eq:eom_alpha}] and [\ref{eq:eom_alpha2}] to first order in $\delta \alpha$ and $\delta R$ about these fixed points yields 
\begin{equation}
\begin{pmatrix}
\delta \dot{R} \\ \delta \dot{\alpha} 
\end{pmatrix}
=\begin{pmatrix}
0 & \mp 1 \\ \pm 2 \kappa_0 & 0
\end{pmatrix}\begin{pmatrix}\delta {R} \\ \delta {\alpha} 
\end{pmatrix}
\end{equation}
of which upon diagonalization leads to the purely imaginary eigenvalue pair $\lambda_{o,\pm}=\pm \sqrt{2 \kappa_0} i$ with corresponding eigenvectors $\vec{x}_{o,\pm}= \frac{1}{\sqrt{2\kappa_0+1}}\left[\pm i ,  \sqrt{2 \kappa_0} \right]^T$, indicating that the fixed points are {\it marginally stable}, and that trajectories close to the fixed points are on center manifolds, and would thus be non-circular closed orbits in the $R$-$\theta$ plane. 
\begin{figure}[tp!]
	\includegraphics[width=1\columnwidth]{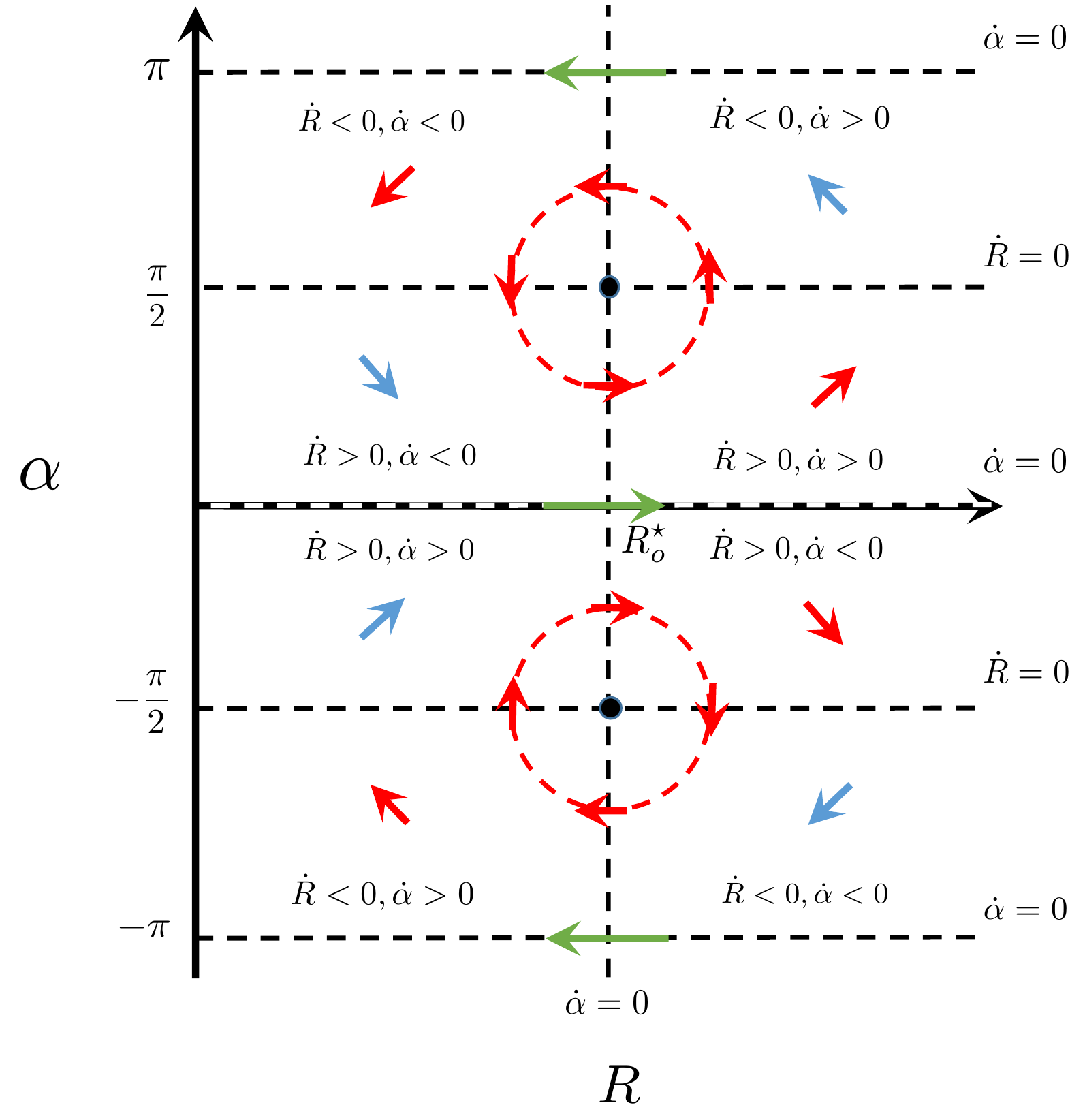}	
	\caption{Phase space of the overdamped system in $R$-$\alpha$ plane. Green arrows represent regions of phase space where trajectories are unstable ($R \to 0$ or $R \to \infty$), red arrows represent confined stable trajectories where circles in particular represent the closed stable trajectories that are predicted by the linear analysis, while blue arrows indicate regions where the stability properties are indeterminate from simply analyzing the time derivatives.}
	\label{fig:phase_space_overdamped}
\end{figure}

In fact, an examination of the dynamical flows in phase space (see Fig. \ref{fig:phase_space_overdamped}) indicates that all initial conditions where $\alpha_{int} \neq \pm \pi/2$ result in confined trajectories where the asymptotic value of $R$ is neither infinity nor zero. This can be seen by noticing that no initial condition other than $\alpha_{int}= \pm \pi/2$ can satisfy the transcendental solution (\ref{eq:transcedental}) for $\alpha= \pm \pi/2$. Similarly $a_{int}=0$ or $\pm \pi$ are the only allowed initial conditions for $R=0$. 

In other words, no trajectories from the regions of phase space for which the subsequent dynamics is indeterminate, indicated by the blue arrows will ever flow into the divergent subspace denoted by the green arrows in Fig. \ref{fig:phase_space_overdamped}. Thus, while points in phase space where $ \alpha=n \pi /2$ results in non-confined trajectories, these form isolated subspaces (only on a line) in phase space that are not accessible from other domains. This also implies that the sign of $\alpha$ and the rototactic active torque is preserved throughout the entire time evolution in overdamped dynamics of this model. 

It then follows that the overdamped active swimmer will always exhibit negative rototaxis in real physical systems where noise is inevitably present, since any finite perturbation away from the unstable line domains ($\alpha = n \pi$), results in stable localized solutions. Note that this negatively rototactic behavior in the overdamped limit stems from the current context of a bottom heavy particle, i.e. $m_1 > m_2$ ,such that $\tilde{\mu} > 0$. A top heavy particle ($m_1 < m_2$) would instead result in positively rototactic behavior. Moreover, note that this argument is also based solely on the dynamical flows and is entirely general in that it is made independent of any linear approximation and is thus valid for arbitrarily far regions in phase space away from the fixed points. 

Furthermore, we also note that the ideal rheotactic solution where the swim orientation is perfectly aligned with the local flow, corresponds in our picture, to an ideal negative rototaxis given by the fixed point solutions $ \left( R_o^{\star}, \alpha_{o,\pm}^{\star} \right)$. On the contrary, ideal positive rototaxis occurs for the unstable line domains $\alpha = \pm \pi$, where swim orientation is always orthogonal to the direction of local flow, for which there is no equivalent counterpart in the rheotactic perspective. As such, rototaxis can be seen as a generalization of rheotactic principles. 

\begin{figure}[tp!]
	\includegraphics[width=\columnwidth]{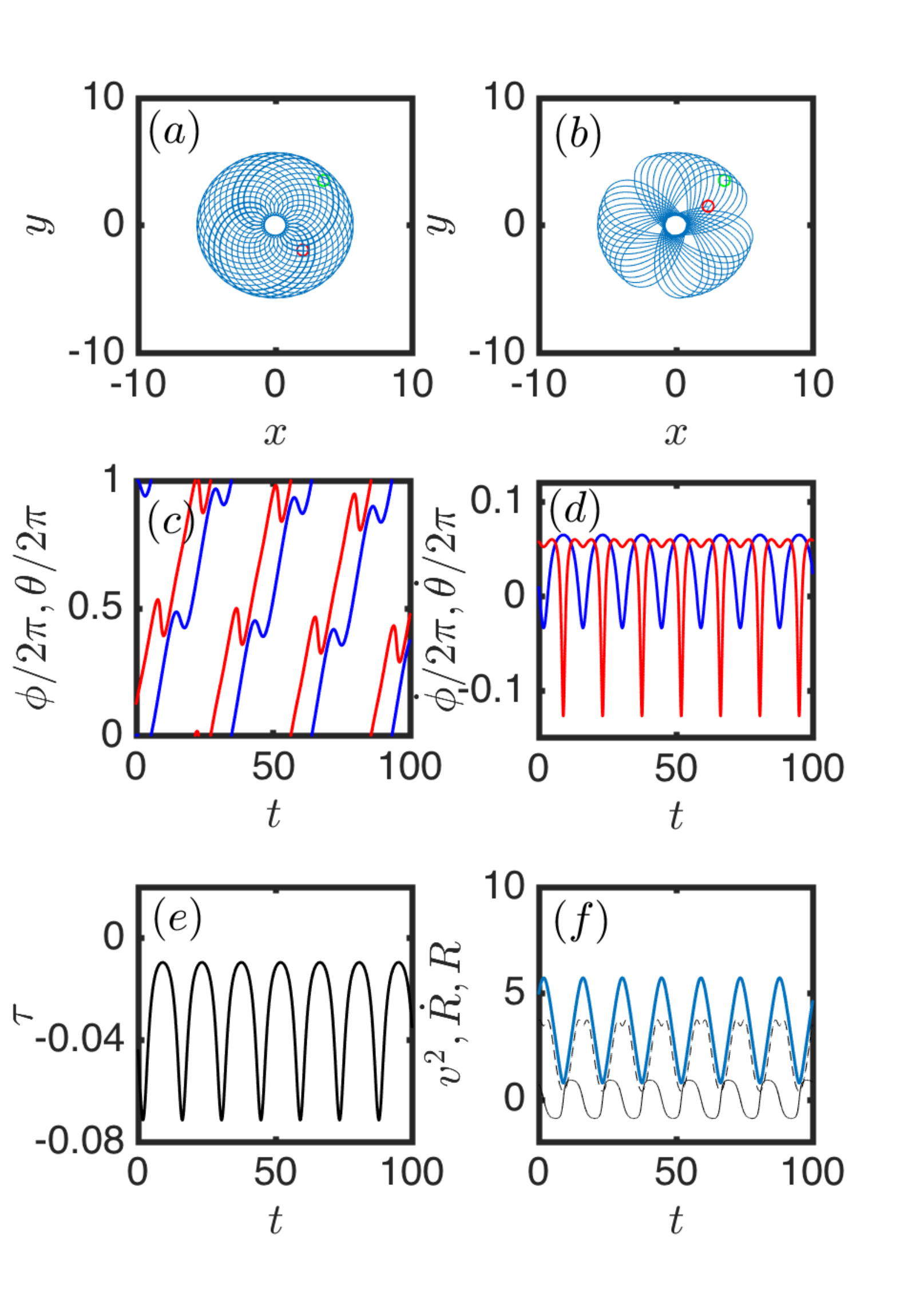}
	\caption{Overdamped dynamics of rototactic active microswimmer.  Rosette and epicyclic dynamical trajectories in the (a) rest and (b) rotating frames in real space. Green and red open circles indicate the initial and final positions of the swimmer in this particular simulation. (c) Orientational $\phi$ (blue) and positional $\theta$ (red) phase and (d) their respective phases velocities. (e) Rototactic torque induced. (f) Rate of change of the radial position $\dot{R}$ (black-solid line), kinetic energy $v^2$ (black-dashed line), and radial distance $R$ from rotation center (blue line). $ \omega^{-1}, l$ and $\frac{m}{2}\tilde{\mu} l \omega^2$ are units of time, length and force respectively.}
	\label{fig:overdamped}
\end{figure}

In the remainder of this section, we focus our attention on the dynamical behavior of the linear response regime in the neighborhood of the fixed points. Explicitly the general solution of the linearized system for some initial conditions $\delta \alpha_0$ and $\delta R_o$ in the neighborhood of $R^{\star}_o$ and $\alpha^{\star}_{o,+}$ is given by:
\begin{equation}
\begin{pmatrix}
\delta R \\ \delta \alpha
\end{pmatrix}
=\frac{1}{\sqrt{2 \kappa_0+1}}\left(A_{+} e^{\sqrt{2 \kappa_0}i t} \begin{pmatrix} i \\  \sqrt{2 \kappa_0} \end{pmatrix}+ A_{-}e^{-\sqrt{2\kappa_0}i t} \begin{pmatrix} -i \\  \sqrt{2 \kappa_0} \end{pmatrix} \right)
\end{equation}
where the coefficients satisfy:
\begin{equation}
A_{\pm}=\frac{\sqrt{2 \kappa_0+1}}{2}\left(  \frac{1}{ \sqrt{2\kappa_0}} \delta \alpha_0 \mp i \delta R_0 \right)
\end{equation}
Note that $A_{+}=A_{-}  $ and $A_{+}=- A_{-} $ in the special cases where $\delta R_0=0$ and $\delta \alpha_0= 0$ respectively. Now, by representing the complex coefficients in polar form as $A_{\pm}=r_A e^{ \pm i \xi}$, we obtain an analytical expression for the dynamical trajectories of the active microswimmer in the linear regime.
\begin{equation}
\begin{pmatrix}
\delta R \\ \delta \alpha
\end{pmatrix} = \delta \tilde{R}_{max} \begin{pmatrix} - \sin \left[(\xi+\sqrt{2 \kappa_0 })t \right] \\  \sqrt{2 \kappa_0} \cos \left[ (\xi+\sqrt{2 \kappa_0}) t \right] \end{pmatrix} 
\end{equation}
where $\xi= \tan^{-1} \sqrt{2 \kappa_0} \delta R_0 / \delta \alpha_0 $ and 
\begin{equation}
\delta \tilde{R}_{max}=\sqrt{\frac{\delta \alpha_0^2}{2 \kappa_0}+\delta R_0^2}
\end{equation}
This expression corresponds geometrically to ellipses centered about $(R_o^{\star},\alpha_o^{\star})$ in the $\alpha$-$R$ plane, with semi-major and minor axes given by $\delta \tilde{R}_{max}$ and $ \sqrt{2\kappa_0}\delta \tilde{R}_{max}$ respectively.

In real space, the trajectories of $\vec{R}(t)$ for dynamics near the fixed points are thus epicyclic in nature characterized by the global rotation frequency $\omega$ and the epicyclic frequency $\Omega_{\xi}=\sqrt{2 \kappa_0}+ \xi$ about the circular solution $R_o^{\star}$. 

The geometrical intuition of this trajectory can be understood as the trace of a particle on an elliptical orbit around a point that itself is rotating about the global rotation center $\hat{O}$ in a circular fashion. This is analogous to the trace of our moon in the rest frame of the sun. Given this geometrical interpretation, $\delta \tilde{R}_{max}$ also represents the maximum deviation in distance of a particle away from $ \left( R_o^{\star}, \alpha_{o,\pm}^{\star} \right)$ at any point along the trajectory, thus explaining the choice of notation we have adopted. 

Direct numerical integration of the equations of motion eqs. [\ref{eq:eom_o}] and [\ref{eq:eom_o2}] also suggest that solutions are in general, closed epicyclical periodic orbits, for non-vanishing torque $\tau(t)$ resulting in rosette-like patterns in both rest and rotating frames of the system. We show representative plots of the trajectories and time evolution of relevant quantities for a specific example with initial conditions $R_{int}=5, \theta_{int}=\pi/4, \phi_{int}=0$ (i.e $\alpha_{int}= \pi/4$)  \cite{overdamped_numerics} (see Fig.\ref{fig:overdamped}). 

Lastly, we also note that while the solutions obtained for the linearized system is only valid strictly in close vicinity of $(R_o^\star,\alpha_o^{\star})$, we find numerically that our results are robust even for points far away from the fixed point in phase space (see Appendix \ref{sec:app}), indicating that the overdamped dynamics of the model rototactic swimmer exhibits negative rototaxis, and in general traces out localized epicyclic-like trajectories.

\section{Inertial Dynamics}
\label{sec:inertia}
\begin{figure*}[htp!]
	\includegraphics[width=0.92\textwidth]{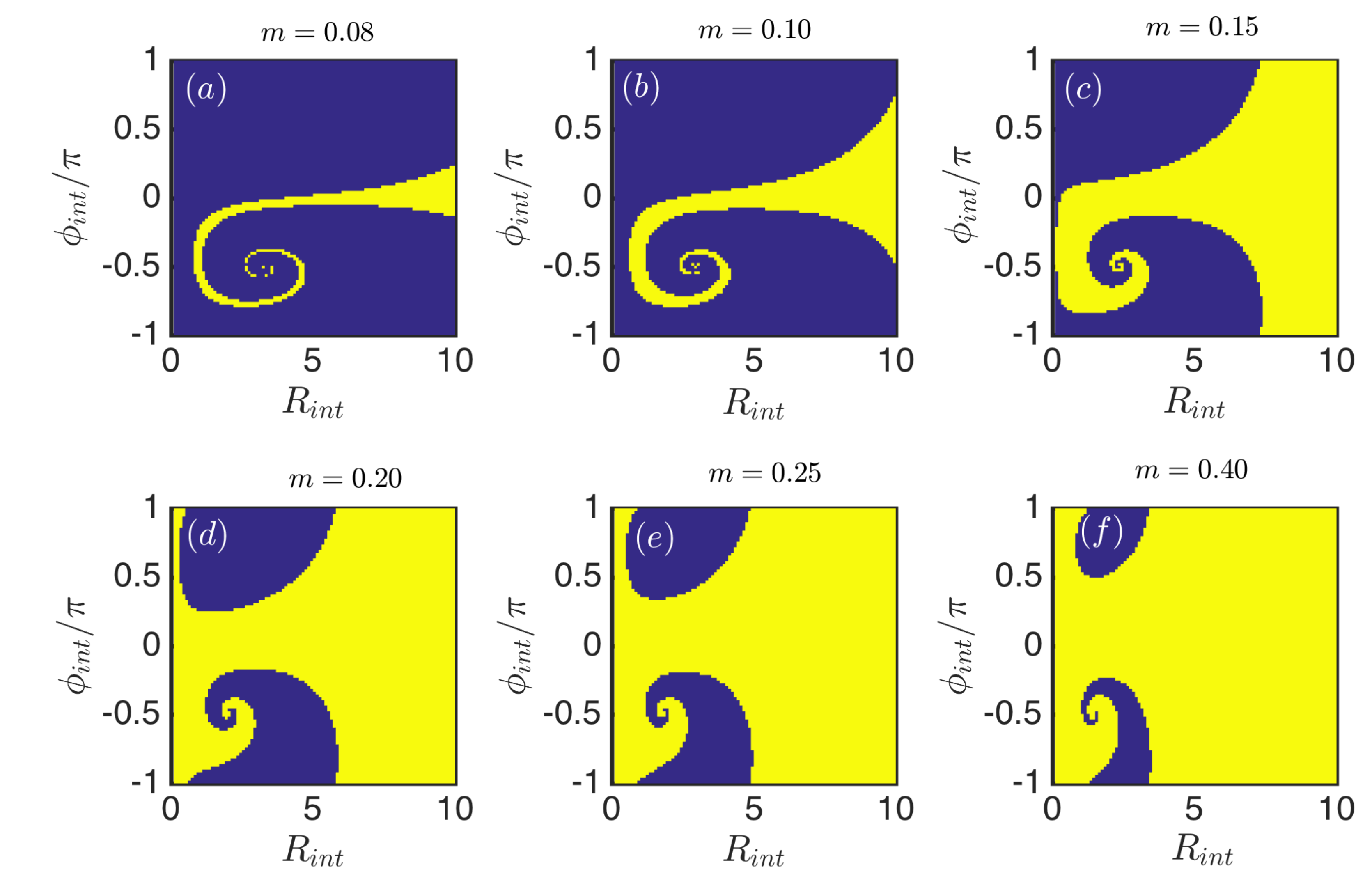}
	\caption{Phase diagram for the inertial dynamics of rototactic particles consisting of symmetrically sized beads of total mass $m$ [(a) 0.08 (b) 0.10 (c) 0.15 (d) 0.20 (e) 0.25 (f) 0.40] for $\gamma_R=0.2$, $\gamma=0.5$ and $\tilde{\mu}=1$. Domains of negative (blue) and positive (yellow) rototaxis corresponding respectively to confined and divergent dynamical trajectories, form complex non-linear patterns over the range of initial conditions $ - \pi < \phi_{int} < \pi $ and $ 0 <R_{int} < 10$.  }
	\label{fig:phase_diagram}
\end{figure*}

In this section we extend our introduction of rototaxis to the general problem where inertia plays a significant role. While existing studies of active matter have predominantly been focused on overdamped systems \cite{ElgetiGompper2015,BechingerVolpe 2016} owing to the typical flow regime where a majority of microswimmers reside, there has been recent interest in understanding the role of inertia \cite{Klotsa2019}. Most notably, experimental work on macroscopic synthetic self-propelled particles are found to exhibit a non-trivial inertial delay between orientation and swim directions that can significantly affect dynamical properties of steady state \cite{ScholzLoewen2018}. In addition, inertial effects can also emerge from unconventional couplings between rotational and translational degrees of freedom \cite{FouxonOr2019}, or a consequence of the finite speed at which information can propagate in systems that exhibit collective motion \cite{AttansiViale2014}. 

Hence, to this end of elucidating the role of inertia in rototaxis, we numerically investigate [\ref{eq:eom}] and [\ref{eq:eom2}], under the same simplifying assumptions of $f(t)=0,g(t)=0$ and $M=0$ for a range of physical parameters and initial conditions. In context of rototaxis, we are primarily interested in the $t \to \infty$ behavior of the dynamics and in particular, its localization properties about its rotation center. Thus, the natural dynamical quantity of interest is the long time or steady state behavior of the radial position along trajectory $R_{t\to \infty}$, which we characterize numerically simply by $R(t=T)$ for some finite simulation time $T$. Moreover, the difference in polynomial and exponential dependence of $R(T)$ on $T$ in scenarios of negative and positive rototaxis respectively, leads to high numerical sensitivity and robustness that unambiguously distinct the two rototactic phases even for relatively short timescales $T$.

Now, without any loss of generality, we also assume that the particle is initially positioned along the positive x-axis such that $\theta_{int}=0$, and probe the initial condition space along the orientational phase $\phi_{int}$ and radial distance $R_{int}$. For simplicity, we also adopt a convention where $\gamma v_0$, $l$ and $\omega ^{-1}$ are used respectively as units of force, length and time.

In Fig. \ref{fig:phase_diagram} we simulate \cite{underdamped_numerics} rototactic particles in the limit of $\tilde{\mu} \to 1$ at various values of the combined mass $m$ at values of  $\gamma_R=0.2$ and $\gamma=0.5$ and employ an effective cut-off to the radial distance at time $T$ to distinguish domains of initial condition space where the dynamics exhibit negative (blue) or positive (yellow) rototaxis. We see that, unlike the overdamped regime, inertial dynamics of the rototactic particle is far more complex and crucially, capable of supporting sizeable two-dimensional domains of positive rototaxis in phase space that are robust to fluctuations in initial conditions unlike the one dimensional lines in the overdamped scenario. 

Moreover, we see from the complex tomography of the phase space across cross-sectional values of $m$, the formation of highly non-trivial patterns in the arrangement of these domains, that lead to the existence of both closed finite, and infinite spiral domain walls, that could even appear simultaneously for a single value of $m$. This represents a high level of non-linearity and initial condition dependence of the dynamics. Here we also note that, for a given $\tilde{\mu}$, the limit of $m \to 0$ corresponds to the overdamped dynamics considered in sec. \ref{sec:overdamped}, and indeed, we see that the domain size of the negative rototactic phase increases with decreasing $m$, and the presence of an asymmetry in the occurrence of negative rototaxis that favors smaller $R_{int}$ and $ \phi_{int} > 0$, as we would expect given the result in \cite{Loewen2019a} and the positively defined global rotation.

\begin{figure}[tp!]
	\includegraphics[width=\columnwidth]{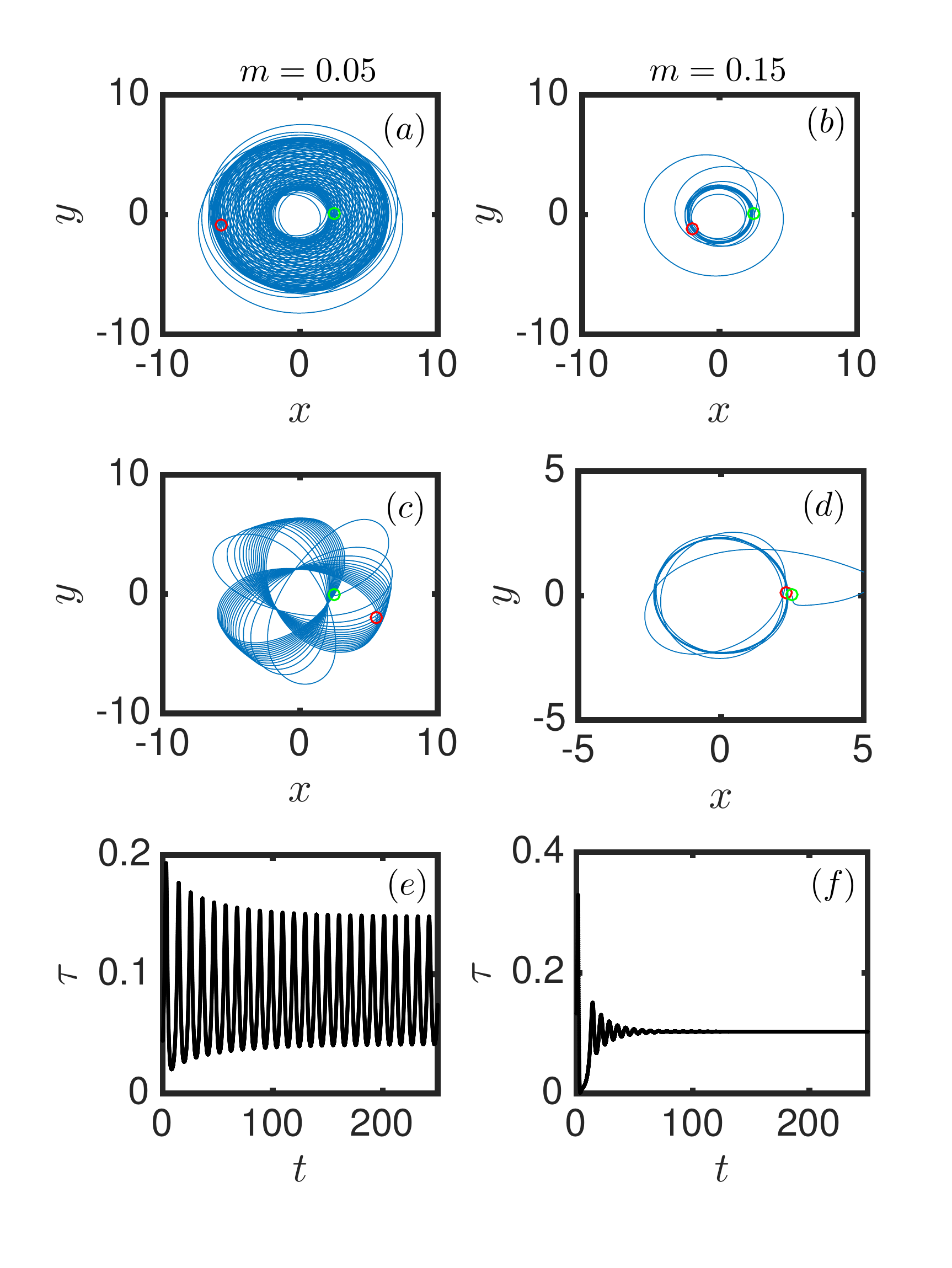}
	\caption{Dynamical behavior of inertial systems for particles of different masses (a,c,e) $m=0.05$ and (b,d,f) $m=0.15$, for otherwise identical system parameters ($\gamma=0.5,\gamma_R=0.25$) and initial conditions  ($\theta_{int}=0,\phi_{int}=\pi/4, R_{int}=2.5$). Epicyclic orbits in the (a) rest and (c) rotating frame reminiscent of overdamped dynamics induced by a (e) periodic oscillating torque and (b) Emergent non-linear circular stable limit cycles in the (b) rest and (d) rotating frames resulting from a (f) constant torque due to phase locking. Green and red open circles represent the initial and final positions of the particle along its trajectory.}
	\label{fig:trajectories}	
\end{figure}

Even more surprisingly, a closer examination of the numerical trajectories reveal that in addition to the epicyclic or rosette like trajectories also found in the overdamped scenario, there exists circular orbits (about the global rotational center) in the steady state that correspond to stable non-linear limit cycles \cite{EinarssonMehlig2014}, which we show examples of in Fig. \ref{fig:trajectories}(c,d). 

These near circular steady state orbits in the rest frame, originate from a rapid damping of oscillations in the time evolution of $\tau$ (see Fig. \ref{fig:trajectories}(f)), such that the steady state torque is a constant. Given that $\tau \sim \sin \alpha$ [\ref{eq:torque}], a constant torque corresponds to stable orbital configurations where the orientational $\theta$ and positional $\phi$ angles are phase locked at some constant difference $\Delta$. This in turn represents a self driven dynamical evolution towards {\it phase synchronization}, that is often found to be emergent in various biophysical models \cite{LevisLiebchen2019,UchidaGolestanian2011}, and is not a solution to the overdamped equations of motion [\ref{eq:eom_o}] and [\ref{eq:eom_o2}] .

In fact, notice that this emergent constant value of $\tau$ in the steady state plays the exact role of an internal torque $M$ that is often placed by hand in various models of circle swimmers that we have thus far neglected, for which the circular orbit is a known solution to the inertial dynamics under a global rotation \cite{Loewen2019a}. It thus follows that the transition between overdamped epicyclic trajectories to circular limit cycles could possibly be understood as the onset of some instability of the circle solution itself. 

To explore this idea in more detail, we write down the equations of motion [\ref{eq:eom}] in polar form \cite{polar}
\begin{multline}
\label{eq:eom_i_polar}
m\left[(\ddot{R}-R\dot{\theta}^2)  \hat{e}_R + (2\dot{R}\dot{\theta} + R \ddot{\theta}) \hat{e}_{\theta}\right] + \gamma \left[R \dot{\theta} \hat{e}_{\theta} + \dot{R} \hat{e}_R\right] -\gamma \omega R \hat{e}_{\theta} 
\\
= \gamma v_0 \cos({\theta - \phi}) \hat{e}_R - \gamma v_0 \sin ({\theta - \phi}) \hat{e}_\theta
\end{multline}
\begin{equation}
\label{eq:eom_i_polar2}
\frac{m}{4} l^2 \ddot{\phi} + \gamma_R (\dot{\phi}- \omega) = - \frac{m}{2} \tilde{\mu} l \omega^2 R \sin (\theta - \phi)
\end{equation}
and assume that the limit cycle (circle of radius $R_0$ about the center of rotation) to be a stationary solution i.e. $\ddot{R}_0 = \dot{R}_0 = 0$ and $\ddot{\theta}_0=0$ to the polar equations of motion eqs. [\ref{eq:eom_i_polar}] and [\ref{eq:eom_i_polar2}] satisfied by $R(t)=R_0$, $\dot{\theta}_0 = \dot{\phi}_0= \omega_0$ and $\theta_0 - \phi_0 \equiv \Delta $ for steady state frequency $\omega_0$ and constant phase difference $\Delta$.

Substitution of this circle solution to [\ref{eq:eom_i_polar}] yields 
\begin{align}
-mR_0 \omega^2_0 &= \gamma v_0\cos \Delta
\\ \gamma R_0(\omega_0-\omega) &= - \gamma v_0 \sin \Delta
\end{align}
and an expression for the radius of the limit cycle
\begin{equation}
R_0 =\frac{\gamma v_0}{\sqrt{m^2 \omega_0^4 +\gamma^2(\omega-\omega_0)^2}} \equiv b \label{eq:radius_limit}
\end{equation} 

Note that these equations imply that the periodic circular trajectory is {\emph {not}} a solution to the equations of motion in the absence of activity (because the terms $\cos \Delta$ and $\sin \Delta$ vanishes for $v_0=0$, such that there is in general no permissible non-zero value of $R_0$), and thus the dynamics discussed here cannot be achieved by a passive rotor. 

\begin{figure*}[htp!]
	\includegraphics[width=\textwidth]{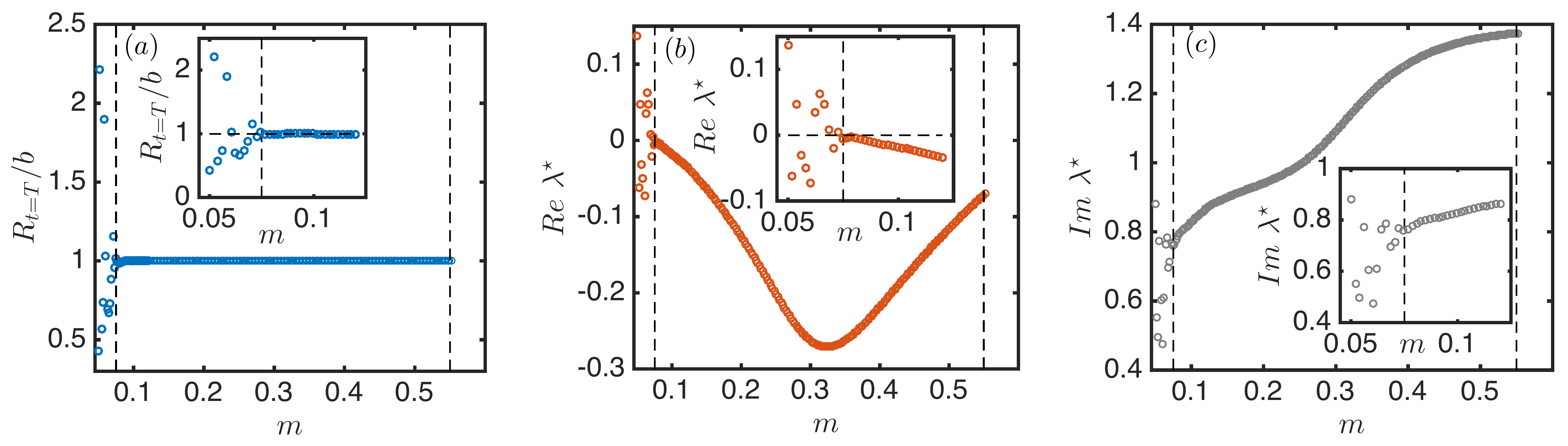}
	\caption{(a) Ratio of the radial position $R_{t=T}$  at time $T$ to the radius of the circular ansatz $b$ as a function of particle mass $m$. (b) Real and (c) imaginary parts of the eigenvalue $\lambda^{\star}$. Inset shows magnified plots near the instability transition of the respective quantities. Note that $R_{t=T}$, and $\lambda^{\star}$ are out of their respective y-axis limits for larger values of $m$, as the particle exponentially escapes the center of rotation.}
	\label{fig:stability}
\end{figure*}

For a system where the actual dynamics is indeed that of the circular limit cycle where $\tau$ approaches a non-zero constant, the radial position $R_{t=T}$ for a sufficiently large $T$ would be exactly equal to the radius $b$, such that $R_{t=T}/b =1$. However as $m \to 0$, the system becomes increasingly overdamped, such that at some point the circle solution becomes unstable ($R_{t=T}/b \neq 1$) although trajectories remain localized. We see this in effect in Fig. \ref{fig:stability}(a), where $R_{t=T}/b$ is strictly one at intermediate values of $m$ but discontinuously becomes ``pseudo-random" at some smaller value of $m$, indicating the presence of an instability transition at some $m=m^{\star}$. Here pseudo-random refers to the behavior where $R_{t=T}$, are seemingly randomly distributed for simulations with arbitrary duration $T$, in contrast to the specific fixed value of $b$ for circular orbits.

To understand the stability properties of the circle solution at the point of transition, we now numerically find the eigenmodes of a linearized system as $m \to 0$ keeping all other parameters and initial conditions constant. Specifically, we consider a (linear) perturbation ($\vec{R}= \vec{R}_0 + \delta \vec{R}$ and $ \phi = \phi_0 +\delta \phi $) to this stationary solution, retaining only contributions to lowest order in $\delta \vec{R}$ and $\delta \phi$. This results in a linearized system of equations in our chosen units of force, length and time (i.e $\gamma v_0= l= \omega ^{-1}=1)$ [\ref{eq:linearized_eom}].

\begin{figure*}[htp!]
	\begin{equation}
	\label{eq:linearized_eom}
	\begin{pmatrix}
	\delta \ddot{R} \\
	\delta \ddot{\theta} \\
	\delta \ddot{\phi} \\
	\end{pmatrix}
	=
	\begin{pmatrix}
	-\gamma/m & 2R_0 \omega_0 & 0
	\\ - 2 \omega_0/ R_0 & -\gamma/m & 0
	\\ 0 & 0 & - 4 \gamma_R / m
	\end{pmatrix}\begin{pmatrix}
	\delta \dot{R} \\
	\delta \dot{\theta} \\
	\delta \dot{\phi} \\
	\end{pmatrix} + 
	\begin{pmatrix}
	\omega_0^2 &  -\sin \Delta / m & \sin \Delta / m 
	\\ - \gamma(\omega_0-1)/mR_0 & - \cos \Delta / mR_0  & \cos \Delta / mR_0 
	\\  - 2 \tilde{\mu} \sin \Delta  &  - 2 \tilde{\mu} R_0\cos \Delta  & 2 \tilde{\mu} R_0\cos \Delta 
	\end{pmatrix}\begin{pmatrix}
	\delta {R} \\	
	\delta {\theta} \\
	\delta {\phi} \\
	\end{pmatrix}
	\end{equation}
\end{figure*}

Upon obtaining the eigenmodes by diagonalizing [\ref{eq:linearized_eom}] numerically, we plot in Fig. \ref{fig:stability}(b,c) the real and imaginary parts of the eigenvalue with the largest real part, which we denote as $\lambda^{\star}$, excluding the zero mode $\lambda_0=0$ with eigenvector $\frac{1}{\sqrt{2}} (\delta \hat{\theta} + \delta \hat{\phi})$. In particular from  Fig. \ref{fig:stability}(b), we see that the behavior of the eigenmodes is such that
\begin{equation}
\lim_{m \to m^{\star}} Re \lambda^{\star} = 0^{-}
\end{equation}
indicating that the fixed point solution of a periodic circular limit cycle becomes unstable at the transition point $m^{\star}$, coinciding with the same point where $R_{t=T}/b$ becomes pseudo-random in Fig. \ref{fig:stability}(a). As such, the transition from overdamped epicyclic like trajectories to stable circular limit cycles is a linear instability transition. While not the focus of our current discussion, we note that our linear instability analysis also correctly identifies the instability transition to the positive rototactic phase, by exhibiting similar discontinuities in the ratio of $R_{t=T}/b$ and $Re \lambda^{\star}$ at a larger value of $m$ in Fig. \ref{fig:stability}(a,b).

\section{Discussion}
\label{sec:discussion}

The localization and aggregation of bacteria in the formation of biofilms are known to be present around vortices in microfluidic vortical flows \cite{YazdiArdekani2012}. However, bacterial shaped (rod-like) particles are at the same time, found to exhibit a preference for alignment to the local flow orientation \cite{BorgninoMehlig2019}. These may at first glance seemingly be phenomena at odds, but under the unifying framework of rototaxis can be understood as two different facets of a single survival strategy, of which one form is preferentially adopted under different circumstances. 

Survival in a rotating environment thus involves strategies that facilitate the rapid adaptation and ease of exchange between two opposing dynamical modes of positive and negative rototaxis. In the previous section, we demonstrated the behavior of a model rototactic swimmer in inertial flow, and showed how inertia can play a role in supporting complex robust domains of both positive and negative rototaxis in phase or parameter space. However, in practice, the mere existence of these domains is in itself insufficient to guarantee the efficient transversal between the two rototactic modes, for it may be energetically unfavorable for the swimmer to re-orientate or position itself, or that it may simply be dynamically unfeasible to do so altogether. 

Thus, it might be a generally dominant strategy for microswimmers to live near the domain walls between the positive and negative rototactic phases, since it would then have to consume far less additional active energy to switch between the two existing rototactic modes. In this context, the highly complex spiral shaped domains in Fig. \ref{fig:phase_diagram} become advantageous for microswimmers since it greatly increases the density of domain walls compared to the finitely bounded phase regions. Moreover, we note that the density of domain walls is maximal near the center of these intertwining spiraling domains, suggesting that the vicinity of these singularities in phase space may be the optimal region for survival in a rotating environment. 

Furthermore, the rototactic principles gleaned from our model may be relevant beyond current context of global rotations. Since the lowest order behavior of a rotation corresponds locally to shear flow, our results and understanding of rototaxis may be relevant to a wide range of microswimmer dynamics and behavior in experimentally relevant shear environments. In fact, it has been recently discovered that transport of bacteria can be heavily suppressed under linear shear \cite{RusconiStocker2014} and even be confined to directional motion out of the shear-flow plane for an oscillatory drive \cite{HopeHaw2016}. Remarkably, it has also been shown, both experimentally and theoretically, that localized cycloid-like periodic trajectories reminiscent to the rosette trajectories found in the negative rototactic phase of our model can also emerge from Poiseuille flow \cite{ZottlStark2013,JunotClement2019}, suggesting that results presented for the global rotation may indeed be relevant to various shear environments beyond linear flow. These phenomena may even prove to provide evolutionary advantages to the bacteria \cite{UppalVural2018}. This connection to shear flow remains to be carefully examined. 

As such, it would be interesting to consider rototaxis in scenarios involving more complex rotational flows with differential vorticity, alternative particle geometries \cite{SchuehHumphries2019,TaylorSalama2019,RusconiStocker2015} and their possible time dependent adaptations, and many body effects in the presence of repulsive and alignment interactions \cite{BialekWalczak2012,HerbertRead2011}.  In even more general flow fields ${\vec u}({\vec r})$ our methods may be useful to study general kinotaxis (alignment along $\nabla {\vec u}^2({\vec r})$ towards the gradient of kinetic flow energy) and enstrotaxis (alignment along $\nabla ({\nabla \times \vec u}({\vec r}))^2$ towards the gradient of the enstrophy).

\section{Conclusion}
\label{sec:conclusion}

In this work, we introduce the idea of rototaxis as a new form of taxis for active particles in rotating environments. We present a model microswimmer capable of generating an active torque that sustains negative rototaxis, epicyclic-like dynamical trajectories localized near its rotation center in the overdamped limit. This minimal bead model is designed to capture essential dynamics of rod-like axial propelled bacteria, but can also easily be extended modularly to more complex geometries and different swim orientations of biological or synthetic swimmers.

In presence of inertia, the negative rototactic phase sustains an additional solution that takes the form of stable circular limit cycles that undergoes a linear instability transition that recovers the rosette trajectories in the overdamped limit. More importantly, inertia also allows for initial conditions where positive rototaxis can occur, where the particle exponentially escapes the rotation center.

These two rototactic phases occupy highly complex regions in phase and parameter space that dramatically increases the density of domain walls in phase space. Living near these domain walls could facilitate the switching in behavior across the two rototactic modes, and may thus prove to be the underlying principle behind viable survival strategies for biological microswimmers in rotating environments.

\acknow{We thank E. Clement, A. Daddi-Moussa-Ider, B. Liebchen, A. M. Menzel and C. Scholz for insightful discussions, and B. Liebchen for suggesting the term ``rototaxis".  H.L. is supported by the German Research Foundation DFG (project LO 418/23-1).}
\showacknow{} 

 
\appendix
\subsection*{Supporting Information (SI)}
\section{Parameter regimes of the overdamped limit}
\label{sec:app_2}

In this appendix, we discuss dynamical regimes and the corresponding range of physical parameters for which the equations of motion and results in the overdamped limit (sec. \ref{sec:overdamped}) are relevant.

First, we rescale the translational equation of motion [\ref{eq:eom}] by introducing $\vec{{R}''} \equiv \vec{R} / l$ and ${t''}=\omega t$, corresponding to the dimensionless variables of position and time respectively, in the limit where $f(t)=g(t)=0$ and $M=0$.
\begin{equation}
\left(\frac{m \omega}{\gamma}\right)\frac{d^2 \vec{{R''}}}{d{t''}^2} + \frac{d\vec{{R''}}}{d{t''}}- \hat{\omega} \times \vec{{R''}}=\left( \frac{v_0}{ \omega l}\right) \hat{n}
\end{equation}
Now, since the terms $d\vec{{R''}}/d{t''}$ and $\hat{\omega} \times \vec{R}''$ are of order $ \sim \mathcal{O}(\vec{{R''}})$, the inertial contribution can be neglected in the limit 
\begin{equation}
\frac{m \omega}{\gamma} \ll \mathcal{O}(\vec{R}'')
\label{eq:translational_inequality}
\end{equation}
and thus the translational equation of motion reduces to the overdamped form [\ref{eq:eom_o}] for regimes where
\begin{equation}
\mathcal{O} (\vec{{R''}}) \sim \frac{v_0}{l \omega}
\label{eq:translational_inequality2}
\end{equation}
Similarly, we obtain by substitution, for the rotational degree of freedom in non-dimensional form
\begin{equation}
\frac{m l^2 \omega}{4 \gamma_R}\ddot{{\phi''}} +\left( \dot{{\phi''}} -1\right) = \frac{m \tilde{\mu} l^2 \omega {R''}}{2 \gamma_R} \sin \left( \theta'' - \phi'' \right)
\end{equation}
Now, we see that the inertial term can be neglected even for a non-vanishing torque for
\begin{equation}
\frac{m l^2 \omega}{4 \gamma_R} \ll 1 \ll {R''}
\end{equation}
Given that the rotational viscosity is typically $\gamma_R \sim \mathcal{O} (\gamma l^2)$, we arrive by use of eqs. [\ref{eq:translational_inequality}] and [\ref{eq:translational_inequality2}] at the limit for the overdamped system
\begin{equation}
\frac{v_0}{l \omega} \gg 1
\end{equation}
This corresponds to a physical picture of strong activity where self-propulsion ($ v_0$) is significantly greater than the typical rotational motion $(l \omega)$.

\section{Robustness of the linear approximation in the overdamped regime}
\label{sec:app}

The dynamical trajectories and solutions of the overdamped microswimmer derived from linear response in the main text is strictly valid only for small neighborhoods around the fixed point. However, numerical simulations seem to suggest that they are in fact good approximations even for initial conditions considerably far from the circular ansatz. 

\begin{figure}[htp!]
	\includegraphics[width=0.98\columnwidth]{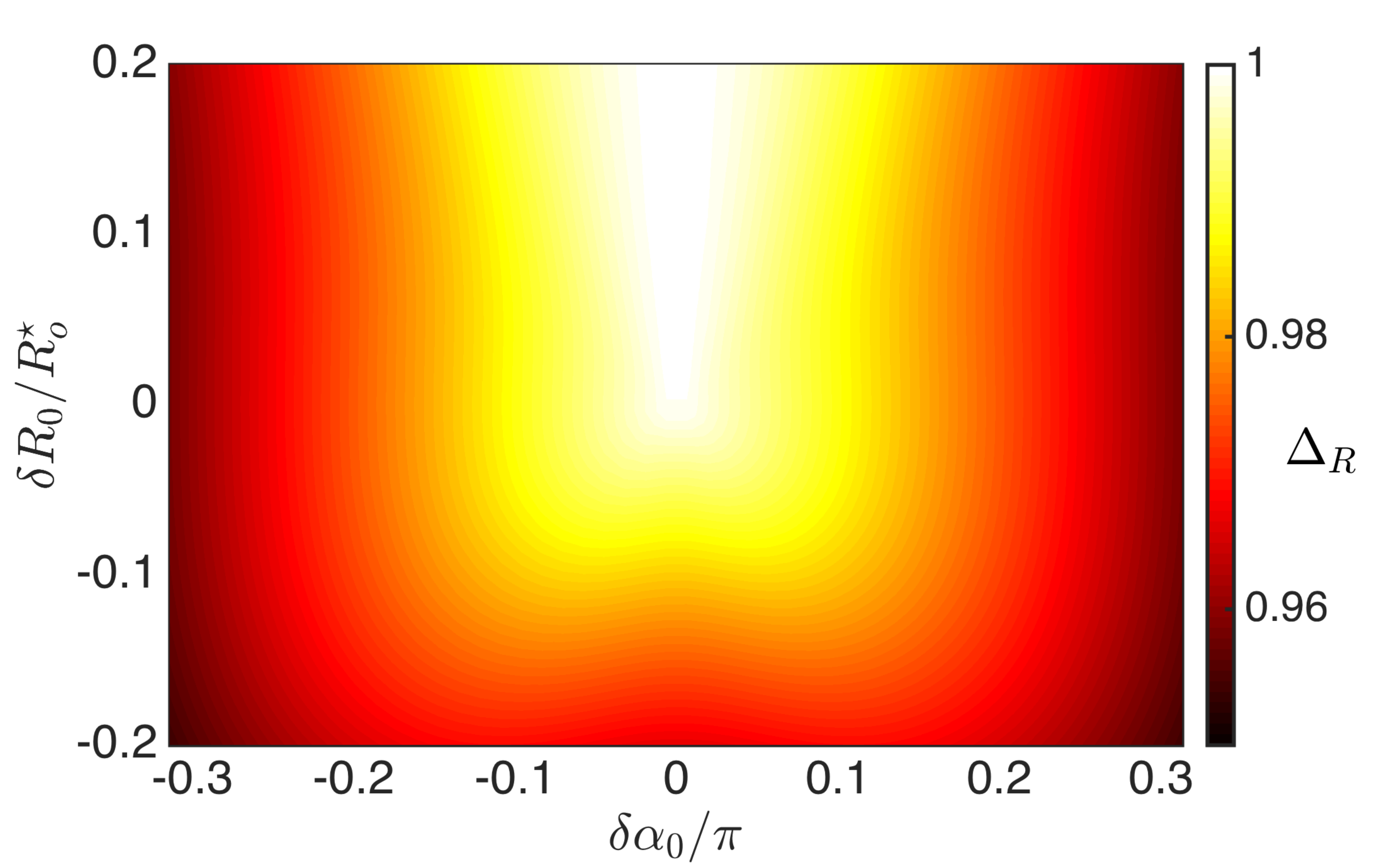}
	\caption{Deviation $\Delta_R \equiv \delta \tilde{R}_{max}/\max_{t} {\vert \vec{R}(t)-\vec{R}^{\star}_o(t)} \vert$, between numerics and linear approximation of the epicyclic amplitude for a range of values of the scaled initial conditions: $\delta R_0/ R^{\star}_o$ and $\delta \alpha_0/\pi$. Parameters of the simulation are $\tilde{\mu}=1, m=1, \gamma_R=2$ and $\alpha_o^\star=\pi/2$, such that $R_o^\star = \sqrt{v_0/ \kappa_0}=2$ in units of $\gamma v_0 ,l$ and $\omega^{-1}$ for force, length and time respectively. }
	\label{fig:r_ratio}
\end{figure}

To illustrate this, we calculate deviations between the maximum radial position (i.e amplitude) relative to the underlying circular ansatz numerically i.e $\max_{t} {\vert \vec{R}(t)-\vec{R}^{\star}_o(t)} \vert $ , and compare that to the analytical prediction $\delta \tilde{R}_{max}$. We find that the actual dynamics of the particle is within $5\%$ of the linear approximation even for initial conditions that are up to $ \pm 10\%$ away from the fixed point, and is thus a reflection of the robustness of our results derived for the linear regime.  See Fig. \ref{fig:r_ratio}, and also note the asymmetry in the $\delta R_0$ axis.


\begin{thebibliography} {200}
\bibitem{overdamped_numerics} Simulation parameters for the overdamped dynamics are $ v_0=20,\gamma=8, \gamma_R=8$ where $ \omega^{-1}, l$ and $\frac{m}{2}\tilde{\mu} l \omega^2$ are units of time, length and force respectively.
\bibitem{underdamped_numerics} Numerical integration of the dynamics is performed using a fourth order Runge-Kutta method with a constant timestep $ \Delta t=10^{-2}$ for a duration of $T= 10^3$. Phase determination is based on a radial cut-off distance of $R_{t=T}=5 \times 10^2$.
\bibitem{polar} Note that the unit vector of the self-propulsion in polar coordinates are obtained by a projection of $\hat{n}$ in the Cartesian basis onto the unit vectors in polar coordinates - $\hat{n} = \cos \phi \hat{e}_x + \sin \phi \hat{e}_y= \cos (\theta - \phi))\hat{e}_R -\sin(\theta - \phi) \hat{e}_\theta$.
\end{thebibliography}
\end{document}